\documentclass[12pt,a4paper]{article}
\input epsf
 \usepackage[dvips]{color}
   \usepackage{multicol}

\begin{document}

\bibliographystyle{apsrev}
\def\half{{1\over 2}}
\def \D {\mbox{D}}
\def\curl {\mbox{curl}\,}
\def \ep {\varepsilon}
\def \lleq {\lower0.9ex\hbox{ $\buildrel < \over \sim$} ~}
\def \ggeq {\lower0.9ex\hbox{ $\buildrel > \over \sim$} ~}
\def\be{\begin{equation}}
\def\ee{\end{equation}}
\def\ber{\begin{eqnarray}}
\def\eer{\end{eqnarray}}
\def \apl {ApJ, }
\def \aps {ApJS, }
\def \pd {Phys. Rev. D, }
\def \prl {Phys. Rev. Lett., }
\def \pl {Phys. Lett., }
\def \np {Nucl. Phys., }
\def\om{\omega}
\def\Om{\Omega_{\rm m}}
\def\la{\lambda}
\def\tla{\Lambda}
\def\tr{\tilde{\rho}}
\def\ra{\rightarrow}\def \l {\Lambda}

\def\apj{{Astroph.\@ J.\ }}
\def\mn{{Mon.\@ Not.\@ Roy.\@ Ast.\@ Soc.\ }}
\def\asta{{Astron.\@ Astrophys.\ }}
\def\aj{{Astron.\@ J.\ }}
\def\prl{{Phys.\@ Rev.\@ Lett.\ }}
\def\pd{{Phys.\@ Rev.\@ D\ }}
\def\nucp{{Nucl.\@ Phys.\ }}
\def\nat{{Nature\ }}
\def\plb {{Phys.\@ Lett.\@ B\ }}
\def \jetpl {JETP Lett.\ }

\begin{center}
\large{{\bf Do recent supernovae Ia observations tend to rule out all the cosmologies?}}
\end{center}

\begin{center}
Ram Gopal Vishwakarma,\footnote{E-mail: rvishwa@mate.reduaz.mx, rvishwak@ictp.it}\\
\emph{Unidad Acad$\acute{e}$mica de Matem$\acute{a}$ticas\\
 Universidad Aut$\acute{o}$noma de Zacatecas\\
 C.P. 98068, Zacatecas, ZAC.\\
 Mexico}\\
\end{center}

\medskip
\begin{abstract}
Dark energy and the accelerated expansion of the universe have been the direct
predictions of the distant supernovae Ia observations which are also supported,
indirectly, by the observations of the CMB anisotropies, gravitational
lensing and the studies of galaxy clusters. Today these results are accommodated 
in what has become the {\it concordance cosmology}: a universe with flat 
spatial sections $t=$ constant with about 70\% of its energy in the form of 
Einstein's cosmological constant $\Lambda$ and about 25\% in the form of dark 
matter (made of perhaps weakly interacting massive particles). Though the
composition is weird, the theory has shown remarkable 
successes at many fronts.

However, we find that as more
and more supernovae Ia are observed, more accurately and towards higher 
redshift, the probability that the data are well explained by the 
cosmological models decreases alarmingly, finally ruling out the concordance
model at more than 95\% confidence level.
 This raises doubts against the 
\emph{`standard candle'}-hypothesis of the supernovae Ia and their use to
constrain the cosmological models. We need a better
understanding of the entire SN Ia phenomenon in order to extract cosmological
consequences from them.

\medskip
\noindent
{\it Subject heading:} SNe Ia: observations, cosmology: theory.


\end{abstract}

\newpage
\noindent
{\bf 1. Introduction}

\noindent
The history of cosmology has probably never witnessed such upheavals as evidenced 
in the past few years, which are primarily due to the observations of distant 
supernovae (SNe) of type Ia. Many colourful exotic models have been 
speculated, which claim to provide satisfactory explanation to  these observation.
It is generally believed that the distant SNe Ia observations predict an 
accelerating expansion of the universe powered by  some hypothetical 
source with negative pressure generally termed as \emph{`dark energy'}.
The simplest and the most favoured candidate of dark energy is a positive 
cosmological constant $\Lambda$, which is though plagued with the horrible 
fine tuning problems. This has led a number of cosmologists to resort 
to  scalar field models of evolving dark energy, which can
produce negative pressure for a potential energy-dominated field
and cause the scale factor to accelerate at late times 
by violating the strong energy condition. 
While the scalar field models enjoy considerable popularity, they have not 
helped us to understand the nature of dark energy at any deeper level. 

Though the idea of the accelerating expansion and dark energy in the framework of 
Einstein's theory is a prediction of the first generation of SNe Ia data, 
however as more and more accurate data get accumulated, the fit to the cosmological
models worsens successively and the recent observations, taken at their 
face value, seem to
rule out all the cosmologies at fairly high confidence levels! This will be 
shown in the present paper. It may be noted that in our goodness-of-fit analysis, we include only those observations which, unlike the SNLS observations, have already 
included the intrinsic dispersion of the SN absolute magnitude in their error 
bars. The SNLS data are not suitable for a goodness-of-fit analysis, as we shall see later.

It should be noted that the precise measurements of the temperature 
anisotropies of the CMB made by the WMAP experiment \cite{Spergel.etal.2003}, 
which appear to offer the most promising determination of the cosmological 
parameters, are often quoted
as providing a direct evidence for an accelerating universe, which is though 
not quite correct. 
The standard interpretation of the WMAP constraints may be misleading, as 
it relies on the assumption of the power law spectrum. 
Blanchard et al. \cite{Blanchard.etal.2003} have shown that the CDM Einstein-de Sitter (EdS)
universe is quite consistent with the WMAP data if the primordial spectrum is
not scale-free.
Hence, taken on their face values, the only apparent prediction made by the WMAP 
observations is a flat geometry, and the decelerating models like the EdS also explain these observations successfully
 \cite{Vishwakarma2003.Blanchard2005}.

\bigskip
\noindent
{\bf 2. $m-z$ relation in Robertson-Walker cosmologies}

\noindent
As the measured quantities of the SNe are the magnitude ($m$) and the 
redshift ($z$),
let us derive, in brief, the usual $m-z$ relation
for SNe Ia in the framework of Einstein's theory, which we shall use for our
analysis. The derivation assumes the simplest model of the universe based on 
the Robertson-Walker (R-W) metric, representing a homogeneous and isotropic 
spacetime. 
Let us assume that the observer at $r=0$ and $t=t_0$ receives light
emitted at $t=t_1$ from a SN of absolute luminosity $L$
located at a coordinate distance $r_1$. The measured (apparent) luminosity $l$
of the SN is defined by 
$l \equiv L/4\pi d _{\mbox{{\scriptsize L}}}^2$, 
where $d_{\rm L}$, the luminosity distance, is given by
\begin{equation}
d _{\mbox{{\scriptsize L}}}=(1+z) S_0 ~ r_1, \label{eq:rone}
\end{equation}
 where $z=S(t_0)/S(t_1)-1$ is the cosmological redshift
of the SN, with $S$ being the R-W scale factor.
(Incidentally the 
luminosities $l$ and $L$ are expressed in terms of
the K-corrected magnitudes $m$ and $M$ as
$l=10^{-2m/5}\times 2.52 \times 10^{-5}$ erg cm$^{-2}$ s$^{-1}$ and 
$L=10^{-2M/5}\times 3.02 \times 10^{35}$ erg s$^{-1}$.) When
written in terms of the magnitudes $m$ and $M$, the above luminosity-redshift relation gets
transformed into the magnitude-redshift relation:
\begin{equation}
m(z;{\cal M}, \Omega_i)={\cal M} +
5 \log\{H_0 d _
{\mbox{{\scriptsize L}}}(z; \Omega_i)
\},\label{eq:mag}
\end{equation}
where ${\cal M} \equiv M - 5 \,\log_{10} H_0 +$ \emph{constant}.
The value of the \emph{constant} depends on the chosen units in which
$d_{\rm L}$ and $H_0$ are measured. For example, if $d_{\rm L}$ is measured
in Mpc and $H_0$ in km s$^{-1}$ Mpc$^{-1}$, then this \emph{constant} comes
out as $\approx$25.
The value of $r_1$ appearing in (\ref{eq:rone}) can be calculated by integrating the R-W metric
for the SN:
\be
r_1 =
\left\{ \begin{array}{ccl}
\vspace{0.4cm}
\sin\left(\frac{1}{S_0} \, \int_0^z \, \frac{{\rm d} z'}{H(z')} \right),& \mbox{when}& k = 1 \\
\vspace{0.4cm}
\frac{1}{S_0} \, \int_0^z \, \frac{{\rm d} z'}{H(z')}, &\mbox{when}& k = 0 \\
\sinh\left(\frac{1}{S_0} \, \int_0^z \, \frac{{\rm d} z'}{H(z')} \right),& \mbox{when}& k = -1.
\end{array}\right. \label{eq:rdist}
\ee
The Hubble parameter $H(z)$, appearing in these
equations is provided by the Einstein field equations.
If the different sources which populate the universe do not interact with each
other and each of them is represented by an equation of state 
$\omega_i\equiv p_i/\rho_i$ (which can be a function of time in general),
 the Friedmann equation then yields
\be
H^2(z) = H^2_0 \, \bigg[ \sum_i \Omega_i \exp\left\{3\int_0^z \frac{1+\omega_i(z')} {1+z'}dz'\right\}
-\Omega_k \, (1 + z)^2 \bigg], \label{eq:hubble}
\ee
where $\Omega_i$ are, as usual, the present day energy densities of the 
different source components in units of the critical density $3H_0^2/8\pi G$
and $\Omega_k \equiv k/S_0^2 H_0^2$ (i denoting non-relativistic matter 
($\rm m$), radiation,($\rm r$), cosmological constant ($\Lambda$), 
quintessence ($\phi$) etc.).
The present value of the scale factor $S_0$, appearing in
equations (\ref{eq:rone}, \ref{eq:rdist}) which measures the present curvature of
spacetime, can now be calculated from
\be
S_0 = H_0^{-1} \sqrt{\frac{k}{
(\sum_i\Omega_i-1)}}.\label{eq:szero}
\ee
We note from equations (\ref{eq:rdist}, \ref{eq:szero}) that the coordinate 
distance $r_1$, and hence $d_{\rm L}$, are sensitive to $\Omega_i$ 
for the distant SNe only. For the nearby SNe ($z<<1$), 
equation (\ref{eq:mag}) reduces to

\begin{equation}
m(z)= {\cal M} + 5 ~\log z,\label{eq:magLowz}
\end{equation}
which can be used to measure ${\cal M}$ by using low-redshift 
supernovae-measurements
(that are far enough into the Hubble flow so that their peculiar
velocities do not contribute significantly to their redshifts).

Now for given ${\cal M}$, $\Omega_i$ and $\omega_i$, these equations can provide the 
predicted value of $m(z)$ at any given 
$z$. We compare this value with the corresponding observed magnitude $m_o$ 
and compute $\chi^2$ from
\be
\chi^2 = \sum_{j = 1}^N \,\left[ \frac{m(z_j;~{\cal M},~\Omega_i,~\omega_i) - m_{o, j}}
{\sigma_{m_{o, j}}}\right]^2,\label{eq:chi}
\ee
where the quantity $\sigma_{m_{o, j}}$ is the uncertainty in the observed
magnitude $m_{o, j}$ of the $j$-th SN.
It may be noted that sometimes the zero-point absolute magnitudes are set arbitrarily 
in different data sets. While fitting the combined data set this situation is 
handled successfully by the constant ${\cal M}$ appearing in equation 
(\ref{eq:chi}), which now plays the role of the normalization constant and
simply gets modified suitably. In this case however it does not represent 
the usual `Hubble constant-free absolute magnitude'
but differs from the latter by an unknown constant (which is though not
needed for the cosmological results). The constant ${\cal M}$ also takes care of the cases
where the data are given in terms of the distance modulus 
$\mu=m(z)-M$, instead of $m$. Equation (\ref{eq:chi}) can also be used in this
case for fitting the data by using $\mu_o$ in place of $m_o$.
 
 Sometimes in the data we are also provided with independent 
uncertainties on some another variable (say, $y$). In this case the 
equation for $\chi^2$ gets modified as
\be
\chi^2 = \sum_{j = 1}^N \,\left[ \frac{\{m(z_j;~{\cal M},~\Omega_i,~\omega_i) 
- m_{o, j}\}^2}
{\sigma_{m_{o, j}}^2+\{\frac{\partial m}{\partial y}(z_j)\}^2 ~\sigma_{y, j}^2}\right],\label{eq:chimod}
\ee
where $\sigma_{y, j}$ is the uncertainty in the observed variable $y$ 
corresponding to the $j$-th SN.

The key point about the SNe Ia data-fitting is that the absolute luminosities $M$
of all the SNe, distant or nearby, are regarded same 
({\it standard candle}-hypothesis). Hence so is the constant ${\cal M}$, 
as it has only one extra parameter
$H_0$ which certainly does not differ from SN to SN. Thus there are two ways
of the actual data fitting: (i) estimate ${\cal M}$ by using equation
(\ref{eq:magLowz}) from the low-redshift SNe, and use this value in equation
(\ref{eq:mag}) to estimate $\Omega_i$ from the high-redshift data alone; 
(ii) use low-, as well as, high-redshift data simultaneously to evaluate all 
the parameters from equation
(\ref{eq:mag}) by keeping ${\cal M}$ as a free parameter. Obviously the second
method gives a better fitting (which we have used throughout this paper).

It is obvious from equation (\ref{eq:chi}) that if the model represents
the data correctly, then the difference between the predicted magnitude and 
the observed one at each data point should be roughly the same size as the 
measurement uncertainties and each data point will contribute to $\chi^2$ 
roughly one, giving the sum roughly equal to
the number of data points $N$ (more correctly $N-$number of fitted parameters
$\equiv$ number of degrees of freedom `dof').
If $\chi^2$ is large, then the fit is bad. However we must quantify our 
judgment and decision about the \emph{goodness-of-fit}, in the absence of 
which, the estimated parameters of the model (and their estimated 
uncertainties) have no meaning
at all. An independent assessment of the goodness-of-fit of the data to the 
model is given in terms of the $\chi^2$-\emph{probability}:  
if the fitted model
provides a typical value of $\chi^2$ as $x$ at $n$ dof, this probability is
given by
\be
P(x, n)=\frac{1}{\Gamma (n/2)}\int_{x/2}^\infty e^{-u}u^{n/2-1} {\rm d}u.
\ee
$P(x, n)$ gives the probability that a model
which does fit the data at $n$ dof, would give a value of $\chi^2$ as large
as $x$ or larger. If $P$ is very small, the model is ruled out.
For example, if we get a $\chi^2=20$ at 5 dof for some model, then the 
hypothesis that {\it the model describes the data genuinely} is  unlikely, 
as the probability $P(20, 5)=0.0012$ is very small.

\bigskip
\noindent
{\bf 3. Fitting cosmological models to different available SNe Ia data sets}

\noindent
Now we have developed enough infrastructure to analyze the observations.
We start our analysis with the data from Perlmutter et al. \cite{Perlmutter.etal.1999}, which 
is one of the important data sets of the first generation
of SN Ia cosmology programs 
\cite{Perlmutter.etal.1999, Perlmutter.etal.1998.Garnavich.etal.1998.Riess.etal.1998.Schmidt.etal.1998}. 
We particularly
focus on the sample of 54 SNe from their `primary fit' C. We have shown
the fit-results in Table 1. The $\Lambda$CDM, as well as the models with the
constant $\omega_\phi$ have a good fit. For example the concordance model
(flat $\Lambda$CDM) has a  $\chi^2=57.7$ at 52 dof with the 
 {\it goodness-of-fit} probability 
$P=$ 27.3\%, representing a good fit. The Einstein de Sitter (EdS) model
($\Omega_{\rm m}=1$, $\Lambda = 0$),
which used to be the favourite model before the SNe Ia observations, 
has a bad fit with $P=$ 0.06\%. In order to take note of the 
   history\footnote{A 
number of theories, e.g., brane cosmology, phantom cosmology, 
Cardassian cosmology, Chaplygin gas cosmology, have been proposed as 
alternatives to the standard paradigm. However
either they reduce to the standard $\Lambda$CDM cosmology
in the present phase of evolution or have even worse fit than the standard
$\Lambda$CDM cosmology . So, they are not included in the fit.}, 
   we have also shown the fit to the 
Bondi-Gold-Hoyle steady state model (for which $m={\cal M}+ 5\log[z\{1+z\}]$)
- the first model which predicted an accelerating universe.

 The existing data points coming from a wide range of different observations
were compiled by Tonry et al. \cite{Tonry.etal.2003}. With many new important additions 
towards higher redshifts, a refined sample of 194 SNe was presented by 
Barris et al. \cite{Barris.etal.2004}. However, the data we are going to consider
next, is the `gold sample' of Riess et al.  \cite{Riess.etal.2004}, which is  
a more reliable set of data with reduced calibration errors arising 
from the systematics. It contains 143 points from the previously published 
data, plus 14 new points with $z>1$ discovered with the Hubble Space Telescope
(HST). While compiling this sample, various points from the previously 
published data were discarded where the 
classification of the supernova was not certain or the photometry was 
incomplete. The fit-results of this data show that although the fits to the 
$\Lambda$CDM and the constant $\omega_\phi$-models are reasonable, they are 
deteriorated
considerably; the probabilities $P$
have reduced to less than
half of the corresponding probabilities obtained in the case of the 
Perlmutter et al' data. Earlier, Roy Choudhury  and Padmanabhan \cite{RoyChoudhury.Padmanabhan2005}
have also shown that the Riess et al' `gold sample' is inconsistent with a 
flat cosmology at 90\% confidence level.

{\small
\begin{center}
\begin{tabular}{lcrrrrrr}
\hline\hline

\multicolumn{1}{c}{Models}&
\multicolumn{1}{c}{$\Omega_{\rm m}$}&
\multicolumn{1}{c}{$\Omega_\Lambda$ or $\Omega_\phi$} &
\multicolumn{1}{c}{$\omega_\phi$} &
\multicolumn{1}{c}{${\cal M}$} &
\multicolumn{1}{c}{$\chi^2$} &
\multicolumn{1}{c}{dof} &
\multicolumn{1}{c}{$P$}\\
\hline
\multicolumn{6}{c}{\textcolor{blue}{\underline{54 SNe from Perlmutter et al. (1999)}}}\\
\vspace{0.1cm}
$\Lambda$CDM (flat) & $0.28\pm0.08$ & $1-\Omega_{\rm m}$ & -1 & $23.94\pm0.05$ & 57.7 & 52 & 0.273\\
\vspace{0.1cm}
$\Lambda$CDM (n.c.)  & $0.79\pm0.47$ & $1.40\pm0.65$ & -1 &$23.91\pm0.06$ & 56.9 & 51 & 0.266\\
\vspace{0.1cm}
cons $\omega_\phi$ (flat) & $0.48\pm0.15$& $1-\Omega_{\rm m}$ & $-2.10\pm1.83$&$23.91\pm0.08$ & 57.2 & 51 & 0.257\\
\vspace{0.1cm}
EdS   & 1 & 0 & &$24.21\pm0.03$ &92.9 & 53 & 0.0006\\
\vspace{0.1cm}
Steady State  & 2 & 0 & &$23.78\pm0.03$ &75.8 & 53 & 0.022\\
\hline
\multicolumn{6}{c}{\textcolor{blue}{\underline{Gold sample of 157 SNe from Riess et al. (2004)}}}\\
\vspace{0.1cm}
$\Lambda$CDM (flat) & $0.31\pm0.04$ & $1-\Omega_{\rm m}$ & -1 &$43.34\pm0.03$& 177.1 & 155 & 0.108\\
\vspace{0.1cm}
$\Lambda$CDM (n.c.)& $0.46\pm0.10$ & $0.98\pm0.19$ & -1 &$43.32\pm0.03$ & 175.0 & 154 & 0.118\\
\vspace{0.1cm}
cons $\omega_\phi$ (flat) & $0.49\pm0.06$ & $1-\Omega_{\rm m}$ & $-2.33\pm1.07$ &$43.30\pm0.04$ & 173.7 & 154 & 0.132\\
\vspace{0.1cm}
EdS  & 1 & 0& &$43.58\pm0.02$ & 324.7 & 156 & $10^{-13}$\\
\vspace{0.1cm}
Steady State  & 2 & 0 & &$43.15\pm0.02$ &318.3 &156 &$10^{-13}$ \\
\hline
\multicolumn{6}{c}{\textcolor{blue}{\underline{164 SNe from Gold $+$ ESSENCE (Krisciunas et al. 2005)}}} \\
\vspace{0.1cm}
$\Lambda$CDM (flat) & $0.30\pm0.04$ & $1-\Omega_{\rm m}$ & -1 &$43.34\pm0.03$ & 190.3 & 162 & 0.064\\
\vspace{0.1cm}
$\Lambda$CDM (n.c.) & $0.48\pm0.10$ & $1.04\pm0.18$ & -1 &$43.32\pm0.03$ & 187.2 & 161& 0.077\\
\vspace{0.1cm}
cons $\omega_\phi$ (flat) & $0.50\pm0.06$ & $1-\Omega_{\rm m}$ & $-2.69\pm1.30$&$43.29\pm0.04$ & 185.5 & 161 & 0.091\\
\vspace{0.1cm}
EdS  & 1 & 0& & $43.60\pm0.02$& 348.2 & 163 & $10^{-15}$\\
\vspace{0.1cm}
Steady State  & 2 & 0 & &$43.15\pm0.02$ &331.1 &163 &$10^{-13}$ \\
\hline
\multicolumn{6}{c}{\textcolor{blue}{\underline{168 SNe from Gold $+$ ESSENCE $+$ 4 SNe from Clocchiatti et al. (2005)}}} \\
\vspace{0.1cm}
$\Lambda$CDM (flat) & $0.29\pm0.04$ & $1-\Omega_{\rm m}$ & -1 &$43.35\pm0.03$ & 200.8 & 166& 0.034\\
\vspace{0.1cm}
$\Lambda$CDM (n.c.)& $0.51\pm0.09$ & $1.12\pm0.16$ & -1 &$43.32\pm0.03$ & 195.8 & 165 & 0.051\\
\vspace{0.1cm}
cons $\omega_\phi$ (flat) & $0.50\pm0.04$& $1-\Omega_{\rm m}$ & $-3.18\pm1.46$&$43.28\pm0.04$ & 193.4 & 165 & 0.065\\
\vspace{0.1cm}
EdS  & 1 &0&&$43.61\pm0.02$ & 367.5 & 167 & $10^{-17}$\\
\vspace{0.1cm}
Steady State  & 2 & 0 & &$43.16\pm0.02$ &338.3 &167 &$10^{-13}$ \\
\hline
\multicolumn{6}{c}{\textcolor{blue}{\underline{Gold sample $+$ 4 SNe from Clocchiatti et al. (2005)}}} \\
\vspace{0.1cm}
$\Lambda$CDM (flat) & $0.30\pm0.04$ & $1-\Omega_{\rm m}$ & -1 & $43.35\pm0.03$& 188.0 & 159& 0.058\\
\vspace{0.1cm}
$\Lambda$CDM (n.c.)& $0.49\pm0.10$ & $1.07\pm0.17$ & -1 &$43.32\pm0.03$& 184.2 & 158 & 0.075\\
\vspace{0.1cm}
cons $\omega_\phi$ (flat) & $0.50\pm0.05$& $1-\Omega_{\rm m}$ & $-2.95\pm1.41$&$43.29\pm0.04$ & 182.1 & 158 & 0.092\\
\vspace{0.1cm}
EdS  & 1 &0&&$43.60\pm0.02$& 345.0 & 160 & $10^{-15}$\\
\vspace{0.1cm}
Steady State  & 2 & 0 & &$43.16\pm0.02$ &325.6 &160 &$10^{-13}$ \\
\hline\hline
\end{tabular}
\end{center}
}
\noindent
{\bf Table 1.} {\small Fits of different cosmologies to available data sets: 
some models have been constrained by the requirement of a flat space 
($\Omega_{\rm m}+ \Omega_\phi=1$), whereas the rest have no constraint 
(n.c.).}

\medskip
We now consider the first results of the ESSENCE project 
\cite{Krisciunas.etal.2005} (made public in August 2005) under which 9 SNe
with redshift in the range 0.5 - 0.8 were discovered jointly with HST and 
Cerro Tololo 4-m telescope. In order to minimize the systematic errors, all
the ground-based photometry was obtained with the same telescope and 
instrument. We consider 7 SNe of this project which have unambiguous redshift
and definite classification, and add them to the `gold sample' resulting in a
reliable sample of 164 SNe. 
The Table 1 shows that the fits to different cosmologies have further worsened
considerably and do not represent a good fit, in any case. Increasing the number
of fitted parameters (for example, in the models with a constant $\omega_\phi$)
improves the fit marginally only.

Next we consider the recent discovery of 5 SNe at redshift $z\approx 0.5$
by the High-$z$ Supernova Search Team \cite{Clocchiatti.etal.2005} (results made
public in October 2005). We consider 4 SNe from this sample for which distances
estimated from the MLCS2k2 (Multi-colour Light Curve Shape) method are 
available, so that we can include them in the previous sample of  
`gold $+$ ESSENCE' which also use the MLCS2k2 method to determine the 
distance moduli. The fit-results of the resulting sample of 168 SNe are
very disappointing. The quality of the fits to different models has 
deteriorated to such an extent that the concordance
model can be rejected at 96.6 \% confidence level! This is an alarming 
situation.
Other models have marginally similar fit and increasing the number of fitted 
parameters does not help significantly. Models with variable $\omega_\phi(t)$
do not help either. For example, if we consider 
$\omega_\phi (z)= \omega_0 + \omega_1 ~z/(1+z)$ with  $\omega_0$, $\omega_1$ 
as constants, we obtain 
$\Omega_{\rm m}=0.42$, $\Omega_\phi=0.42$, $\omega_0=-4.95$ and  
$\omega_1=2.83$ as the best-fitting solution with $\chi^2=193.07$ at 163 dof
and $P=5.4$\% (in fact, the model is very degenerate in this case and the
parameters wander around near the minimum  $\chi^2$ in a flat valley of some 
complicated topology). 

It has been mentioned that the ESSENCE data are affected by the selection
bias which can affect their cosmological use \cite{Krisciunas.etal.2005}. 
However, we do not find any such effect from the fitting, as is clear from 
the Table. We notice that the parameters and their uncertainties
estimated by using the ESSENCE sample are absolutely consistent with those
from the use of gold sample or the gold+Clocchiatti et al.'s sample; and we 
do not find any smoking gun pointing out that this sample is completely
different from the others. However, even if we exclude the ESSENCE sample
from the fit, the concordance model is still ruled out at more than 94\%
confidence level.  
 
Finally we consider the recently published (made public in October 2005) first year-data 
of the planned five-year SuperNova Legacy Survey (SNLS) \cite{Astier.etal.2005}.
However, the SNLS data have been analyzed in a 
different way than the other data sets we considered earlier  and it does not make sense to
add them for a joint analysis. Hence we limit ourselves to commenting on the way SNLS
data have been analysed.

In the SNLS, the authors claim to have achieved high precision from improved statistics
and better control of systematics by using the multi-band rolling
search technique and a single imaging instrument to observe the same fields.
Their data set includes 71 high redshift SNe Ia in the redshift range 0.2 - 1 
from the SNLS, together with 44 low redshift SNe Ia compiled from the 
literature but analyzed in the same manner as the high-z sample.
As the SNLS data have been analyzed differently, the fitting procedure 
followed by the authors is also different. In order to calculate $\chi^2$, 
they use
\be
\chi^2 = \sum_{j = 1}^N \,\left[ \frac{\{\mu(z_j;~{\cal M},~\Omega_i) 
- \mu_{o, j}\}^2}
{\sigma_{\mu_{o, j}}^2+\sigma_{\rm int}^2}\right],\label{eq:chilegacy}
\ee
where $\sigma_{\rm int}$, is the (unknown) intrinsic dispersion of the 
SN absolute magnitude which, {\it unlike the other data sets}\footnote
{It should be noted that all the other observations, we have considered before,
already include the intrinsic dispersion of the 
SN absolute magnitude in their error bars $\sigma_{m_o}$ or  $\sigma_{\mu_o}$,
which has been estimated by reasonable methods}, is not included in 
the $\sigma_{\mu_o}$; rather it has been used as an adjustable free parameter 
to obtain $\chi^2$/dof = 1.
 We shall return to this issue later for our comments.
First we want to verify if we can reproduce the results of Astier et al. from 
equation (\ref{eq:chilegacy}) by using their
calculated $\mu_o$ (given in columns 7 of their Tables 8 and 9) instead of
using their `stretch' and `color' parameters, which do not seem necessary once 
we have $\mu_o$. We find that by fixing $\sigma_{\rm int}=0.13$,
we get $\Omega_{\rm m}=1-\Omega_\Lambda=0.26$, ${\cal M}=43.16$
with $\chi^2$/dof = 1.00; and
$\Omega_{\rm m}=0.31$, $\Omega_\Lambda=0.81$, ${\cal M}=43.15$
with $\chi^2$/dof = 1.01. This is exactly what Astier et al have obtained.

The intrinsic dispersion in the absolute magnitude of SN Ia (combined with dust extinction of the host galaxy) can be estimated only statistically (unlike the photometric error, which can be estimated from the photometry of the individual SN Ia). Unfortunately we do not have a reliable way to estimate dust extinction or pure intrinsic dispersion of SNe Ia separately.  
However, the introduction of $\sigma_{\rm int}$ in equation 
(\ref{eq:chilegacy}) is justified only when we use independent measurement
uncertainties $\sigma_{\rm int, j}$ on the parameter (as we have mentioned 
earlier in equation (\ref{eq:chimod})), instead of using it as a free 
parameter.
The latter case is just equivalent to increasing the error bars suitably in 
order
to have a desired fit. In this way one can fit any model to the data. For
example, the EdS model can also have a similar
fit by considering $\sigma_{\rm int}=0.258$: ${\cal M}=43.46$ with
$\chi^2$/dof = 1.00. 
This shows that the approach does not have any predictive power. 
One could choose the variable $\omega_\phi(z)$ such that it gives a
lower $\chi^2$/dof with the same $\sigma_{\rm int}=0.13$.
Also, for a similar $\sigma_{\rm int}$, one can obtain a reasonable value
of the reduced $\chi^2$ for another model. For example, increasing
$\sigma_{\rm int}$ only to $\sigma_{\rm int}=0.16$, the model $\Omega_{\rm m}=0$, $\Omega_\phi=0$ gives $\chi^2$/dof =0.86. However, one cannot physically 
test the viability of the model so that one can take any 
decision. This happens because the present approach 
(which simply assumes, rather than tests, that the data have a good fit to the model) 
 prohibits an independent assessment
of the goodness-of-fit-probability $P$, in the absence of which the estimated
parameters do not have any significance.
All one can do, with the present approach, is that one can compare the 
goodness-of-fit of different models.
For example, with $\sigma_{\rm int}=0.13$, the EdS model has a worse fit
than the $\Lambda$CDM model, giving  $\chi^2$/dof = 2.7. In this context, it
would be remarkable that the results of the SNLS data are consistent with a flat geometry whereas the best-fit values of the other observations give a larger value of the $\Omega_{\rm total}$.

\bigskip
\noindent
{\bf 4. Conclusion}

\noindent
Supernovae Ia observations have profoundly changed cosmology
by predicting an accelerated rate of cosmic expansion, and thus a repulsive 
dark energy component - an issue which is regarded as an almost mature science
now.
However, as more and more accurate data get accumulated, thanks to the
remarkable progress made in various types of astrophysical
and cosmological observations in recent years, they do not seem to fit
any cosmology. The recent 
observations, taken at their face values, seem to rule out all the cosmologies at fairly high confidence levels. 
Though these probabilities may not be regarded sufficient to rule out the models 
completely, however, they are high enough to point out towards the alarming 
trend of the recent data: as you add newer data to the older samples,
the goodness-of-fit-probability from the resulting samples successively decreases. 
Though the fit improves in some cases if we do not stick to the concordance 
model, however, this is 
inconsistent with the anisotropy measurements of CMB 
which predict a flat space.

The recently made SNLS  
observations by the Supernova Legacy Survey \cite{Astier.etal.2005}
are analyzed in a different way. While the other observations estimate the 
intrinsic scatter in the absolute magnitude $\sigma_{\rm int}$ from the 
nearby data, the SNLS estimates it from all the observed data. For this purpose, $\sigma_{\rm int}$ is considered as a free parameter to be estimated from all the data by requiring that it gives a good fit. Obviously, this data  
is not suitable for a goodness-of-fit analysis. 
It must be noted that our result (that the recent 
observations seem to rule out all the cosmologies at fairly high confidence levels) 
is deduced from those observations only which, unlike the SNLS data, have already 
included $\sigma_{\rm int}$ in their error 
bars. 

Assuming that the standard big bang cosmology is correct, the present situation
is pointing out towards some flaws in our understanding of the SN Ia 
phenomenon and towards the futility of the use of SNe Ia in order to constrain
cosmological models. We need better understanding of the entire SN Ia 
phenomenon in 
order to test the empirical calibrations that are so confidently extrapolated 
at high redshifts. Similar conclusions have also been drawn by 
Clocchiatti et al. \cite{Clocchiatti.etal.2005} from a smaller sample of data. However, this is more
evident from the present analysis of a bigger sample of data. This view is also supported by
Middleditch \cite{middleditch} who argues that SNe Ia seem to be affected by some systematic 
effects which alone, without invoking any dark energy, could make them too faint for their redshifts.
It is argued that it may be impossible to get a clean sample of SNe Ia which are free from this 
kind of effects \cite{noclean}.

This reminds us of a similar story of the $m-z$ test for
galaxies in 1970s - it was agreed upon finally that uncertain evolutionary 
effects tend to dominate at high redshifts. 
Though most studies confirm that the luminosity properties of SNe Ia at 
different
 redshift and environments are similar \cite{Perlmutter.etal.1999, Sullivan.etal.2003}, however, there are other theoretical studies which have found
variations indicating evolutionary effects \cite{Dominguez.etal.1998.Hoflich.etal.1998}. It has been shown by Drell et al. \cite{Drell.etal.2000} that 
the peak luminosities estimated for individual SNe Ia by two 
different methods are not entirely consistent with one another at high 
redshifts, $z \sim 0.5$. 
If evolution was entirely absent, the differences between them should
not depend on redshift. They further showed that the three luminosity 
estimators in practice
(the multicolor light curve shape method, the template fitting method, and
the stretch factor method) reduce the dispersion of distance moduli about
best fit models at low redshift, but they do not at high redshift, indicating
that the SNe have evolved with redshift. This view is also corroborated by some
recent observations which may go against the `standard candle'-hypothesis of
SNe of type Ia, we mention the following two:

\begin{enumerate}\item
 SNLS-03D3bb, which is a recently observed high redshift ($z=0.244$) type Ia
SN with extreme unusual features and no
obvious analogue at low redshifts. It does not obey the usual lightcurve
shape-luminosity relationship for SNe Ia that allows them to be calibrated as 
standard candles \cite{SNLS-03D3bb}. 

\item
 Observations of two supernova remnants DEM L238 and DEM L249 made with the {\it Chandra} and {\it XMM-Newton} X-ray satellites in the {\it large Magellanic cloud} \cite{snr} 
may also be mentioned. While the  presence of Fe-rich gas at the centres of 
these objects suggests that they are remnants of type Ia supernova 
explosions, the standard model of type Ia supernova remnants cannot explain the presence
of relatively dense supernova ejecta with long ionization timescales.
\end{enumerate}

We may recall that in the 1970s-80s astronomers tried to fit number counts of 
radio sources to cosmological models and improved the fit by adding
multi-parameter evolutionary functions like luminosity evolution, number
density evolution, etc. This gave interesting possibilities of evolution
though  had no predictive value for the model. One can do similar exercise
to improve the fit of the SNe Ia, particularly by keeping the
dark energy equation of state parameter $\omega_\phi(z)$ completely free as a
function of $z$.

\medskip
\noindent
{\bf Note added in the proof:}
It may be mentioned that a more recently made observation \cite{riess07} claims improvements in the data and provides  better fits to the cosmological models. However, this sample does not include the ESSENCE data \cite{Krisciunas.etal.2005} in the analysis, and hence our analysis is still meaningful.

\bigskip
\noindent
{\bf Acknowledgements} The author thanks Pierre Astier 
for providing necessary clarifications on different issues regarding their
astro-ph/0510447. Thanks are also due to Jayant V. Narlikar and Adam Riess 
for useful comments.
The author acknowledges UAZ, Mexico for funding this 
research through the Promep-research grant and thanks Abdus Salam ICTP for hospitality.

\end{document}